\documentclass[11pt,a4paper]{article}
\pdfoutput=1
\usepackage{jheppub}

\usepackage{bold-extra}
\usepackage{enumerate}
\usepackage{fancyhdr}
\usepackage{lscape}
\usepackage{mathrsfs}
\usepackage{multirow}
\usepackage[final]{pdfpages}
\usepackage{rotating}
\usepackage{subfig}
\usepackage{textcomp}
\usepackage{url}
\usepackage{verbatim}
\usepackage{wrapfig}
\def\nn{\nonumber}

\def\gsim{\mbox{\raisebox{-.6ex}{~$\stackrel{>}{\sim}$~}}}

\title{Semiclassical Approach to Heterogeneous Vacuum Decay}

\author[a]{Benjam\'{i}n Grinstein,}
\author[b]{and Christopher W. Murphy}

\affiliation[a]{Department of Physics, University of California, San Diego, \\9500 Gilman Drive, La Jolla, CA 92093, USA}
\affiliation[b]{Scuola Normale Superiore, \\ Piazza dei Cavalieri 7, Pisa 56126, Italy}

\emailAdd{bgrinstein@ucsd.edu}
\emailAdd{christopher.murphy@sns.it}

\abstract{We derive the decay rate of an unstable phase of a quantum field theory in the presence of an impurity in the thin-wall approximation. This derivation is based on the how the impurity changes the (flat spacetime) geometry relative to case of pure false vacuum. Two examples are given that show how to estimate some of the additional parameters that enter into this heterogeneous decay rate. This formalism is then applied to the Higgs vacuum of the Standard Model (SM), where baryonic matter acts as an impurity in the electroweak Higgs vacuum. We find that the probability for heterogeneous vacuum decay to occur is suppressed with respect to the homogeneous case. That is to say, the conclusions drawn from the homogeneous case are not modified by the inclusion of baryonic matter in the calculation. On the other hand, we show that Beyond the Standard Model physics with a characteristic scale comparable to the scale that governs the homogeneous decay rate in the SM, can in principle lead to an enhanced decay rate.}

\arxivnumber{1509.05405}

\begin{document}
\maketitle

\section{Introduction}
The discovery of the Higgs boson combined with the absence of any observation of physics beyond the Standard Model (SM) has an intriguing implication for the ultimate fate of the universe. With the measured value of the mass of the Higgs boson as the final SM input, $m_H = 125.09 \pm 0.21 (\text{stat.}) \pm 0.11 (\text{syst.})$ GeV~\cite{Aad:2015zhl}, state of the art calculations suggest that the electroweak (EW) vacuum is unstable~\cite{Degrassi:2012ry, Buttazzo:2013uya, Andreassen:2014gha} (see also references therein). However, the lifetime of the universe is computed to be many, many orders of magnitude larger than the current age of the universe; the EW vacuum is said to be metastable. Due to this metastability, there is no need to invoke beyond the Standard Model (BSM) physics to explain the observed age of the universe. See~\cite{DiLuzio:2015iua} for a recent review.

Much of the intuition developed for calculating the decay rate of an unstable phase of a quantum field theory (QFT) is based on theory developed for systems that can be studied in the laboratory. In fact, the framework developed by Coleman and collaborators~\cite{Coleman:1977py, Callan:1977pt, Coleman:1980aw} for calculating decay rates in QFT, which is what is used to determine whether the Higgs vacuum is metastable or unstable, is the relativistic, four-dimensional analog of the previously developed methods used in statistical mechanics. 

In particular, what was calculated when it was suggested that the universe is metastable is the homogeneous decay rate. However, in most commonly studied systems it the heterogeneous decay rate (rather than homogeneous) that dominates the catalysis of the phase transition. In a typical laboratory system, a phase transition is seeded by the presence of an impurity either in the bulk of the unstable phase, or by the boundary of the system. Since the intuition for the Higgs vacuum decay rate has been developed from these systems, and since the universe is not just a constant electroweak vacuum, it seems worthwhile to study the heterogeneous decay rate of the Higgs vacuum. 

Classical nucleation theory, including the heterogeneous case, seems to have first been introduced by Volmer for liquid nuclei~\cite{Volmer:1929, Volmer:1939}. The extension to the case of crystal nucleation on surfaces was made by Turnbull~\cite{Turnbull:1949a, Turnbull:1949b, Turnbull:1950} (and their respective collaborators). The nucleation theories that most closely resemble Coleman's work were subsequently developed by Langer~\cite{Langer:1967ax, Langer:1969bc}.\footnote{This is likely an incomplete list of references as the subject is rather old. However, as a consolation, see~\cite{AIC:AIC690210502, :/content/aip/journal/jpcrd/14/3/10.1063/1.555734, Caupin20061000} for some reviews of the experimental confirmation of classical nucleation theory and data on the decay rates of superheated liquids.}

There is a relatively small amount of literature on induced vacuum decay in QFT. Refs.~\cite{Affleck:1979px, Voloshin:1993ks} looked at the induced decay rate due to single particles. An analysis based on these methods, Ref.~\cite{Enqvist:1997wv}, was done to see if cosmic rays could catalyze vacuum decay. Classical catalysis, rather than quantum tunneling, due to Hawking radiation has been investigated in Refs.~\cite{Moss:1984zf, Green:2006nv, Cheung:2013sxa}. Black holes as the seeds of bubble nucleation were discussed in Refs.~\cite{Hiscock:1987hn, Berezin:1987ea, Arnold:1989cq, Berezin:1990qs, Gregory:2013hja, Burda:2015isa, Burda:2015yfa}. 

Enhanced vacuum decay rates have been investigated outside the context of an impurity induced decay as well. A general analysis of tunneling in theories with multiple scalar fields was made by Refs.~\cite{Balasubramanian:2010kg, Czech:2011aa} using the thin-wall approximation. The decay rate of the Higgs vacuum at finite temperature was recently updated in Ref.~\cite{Rose:2015lna}. While Refs.~\cite{Espinosa:2007qp, Fairbairn:2014zia, Hook:2014uia, Herranen:2014cua, Shkerin:2015exa, Kearney:2015vba, Espinosa:2015qea} investigated the stability of the Higgs vacuum in the early universe with a particular focus on inflation. 

When studying vacuum decay outside of the SM, it is natural to expect that the lifetime of the electroweak vacuum would in general be different from the prediction of the SM. For example, Refs.~\cite{Branchina:2013jra, Branchina:2014rva, Branchina:2015nda} {color{red}considered} the SM supplemented by Planck-scale suppressed higher-dimensional operators such that the vacuum structure of the theory is different from that of the SM. In was shown in~\cite{Branchina:2013jra, Branchina:2014rva, Branchina:2015nda} that the lifetime of the EW vacuum in these theories can indeed be much shorter than it is in the SM. For additional discussion of this scenario, see~\cite{DiLuzio:2015iua, Espinosa:2015qea}.

The rest of the paper is organized as follows. In Section~\ref{sec:D}, we derive the decay rate of an unstable phase of a quantum field theory in the presence of an impurity in the thin-wall approximation. This derivation is based on the how the impurity changes the (flat spacetime) geometry relative to case of pure false vacuum. Then, in Sec.~\ref{sec:ex}, two examples are given that show how to estimate some of the additional parameters that enter into this heterogeneous decay rate. After that, Sec.~\ref{sec:H} applies the formalism developed in previous two sections to the EW vacuum of the SM. The goal is to determine if the conclusion drawn in the homogeneous analysis about BSM physics not being necessary to explain the observed age of the universe is still valid in the heterogeneous case. We find that baryonic matter (stars), which acts as an impurity in the electroweak (metastable) vacuum, leads to a heterogeneous decay rate that is suppressed with respect to the homogeneous case. That is to say, the conclusions drawn from the homogeneous case are not modified by the inclusion of baryonic matter in the calculation. Additionally, we confirm that BSM physics with a characteristic scale comparable to the scale that governs the homogeneous decay rate in the SM (or very dense physics) can lead to an enhanced decay rate. 

\section{Derivation of the Heterogeneous Decay Rate}
\label{sec:D}
In this section, we derive the decay rate of a metastable phase in the presence of an impurity in the thin-wall approximation. The decay rate can generically be written as,
\begin{equation}
\label{eq:Gam}
\frac{\Gamma}{S} = A\, e^{- B},
\end{equation}
where $S$ is a (in general dimensionful) symmetry factor to be discussed in what follows. In QFT, the coefficient $B$ is equal to the change in the Euclidean action of the system due to the appearance of a bubble of stable phase within the bulk metastable phase, $B = S_E^{\star}$. In the thin-wall approximation, this change in the action is given by,
\begin{equation}
\label{eq:S}
S_{E,1} = - \frac{\pi^2}{2} r_1^4 \epsilon_{13} + 2 \pi^2 r_1^3 S_{13},
\end{equation}
where we are using a notation similar to that of Coleman~\cite{Coleman:1977py} (the subscripts are for later convenience) with $\epsilon_{13}$ and $S_{13}$ being the difference in the energy density and the surface tension between the true and false vacua respectively. $r_1$ is the radius of the bubble. Extremizing the action with respect to the radius yields the critical bubble size needed for nucleation to occur,
\begin{equation}
r_1^{\star} = \frac{3 S_{13}}{\epsilon_{13}}.
\end{equation}
Substituting $r_1^{\star}$ back into $S_E$ yields the critical action,
\begin{equation}
\label{eq:SEho}
S_{E,1}^{\star} = \frac{27 \pi^2 S_{13}^4}{2 \epsilon_{13}^3}.
\end{equation}

In this work, we will only be interested in order of magnitude estimates for the pre-exponential factor, $A$, but it is an important factor to keep track of nonetheless. To be more precise regarding Eq.~\eqref{eq:Gam}, the quantity of interest for a homogeneous QFT is the decay rate per unit volume, $\Gamma / \mathcal{V}_3$. Recall that this is so because the coefficient $B$ is invariant under (Euclidean) spatial translations in this case, leading to the probability for tunneling being proportional to the volume of spacetime. Then, by dimensional analysis, the pre-exponential factor of $\Gamma / \mathcal{V}_3$ must be proportional to $1 / (r_1^{\star})^4$.

\begin{figure}
  \centering
\includegraphics[width=0.5\textwidth]{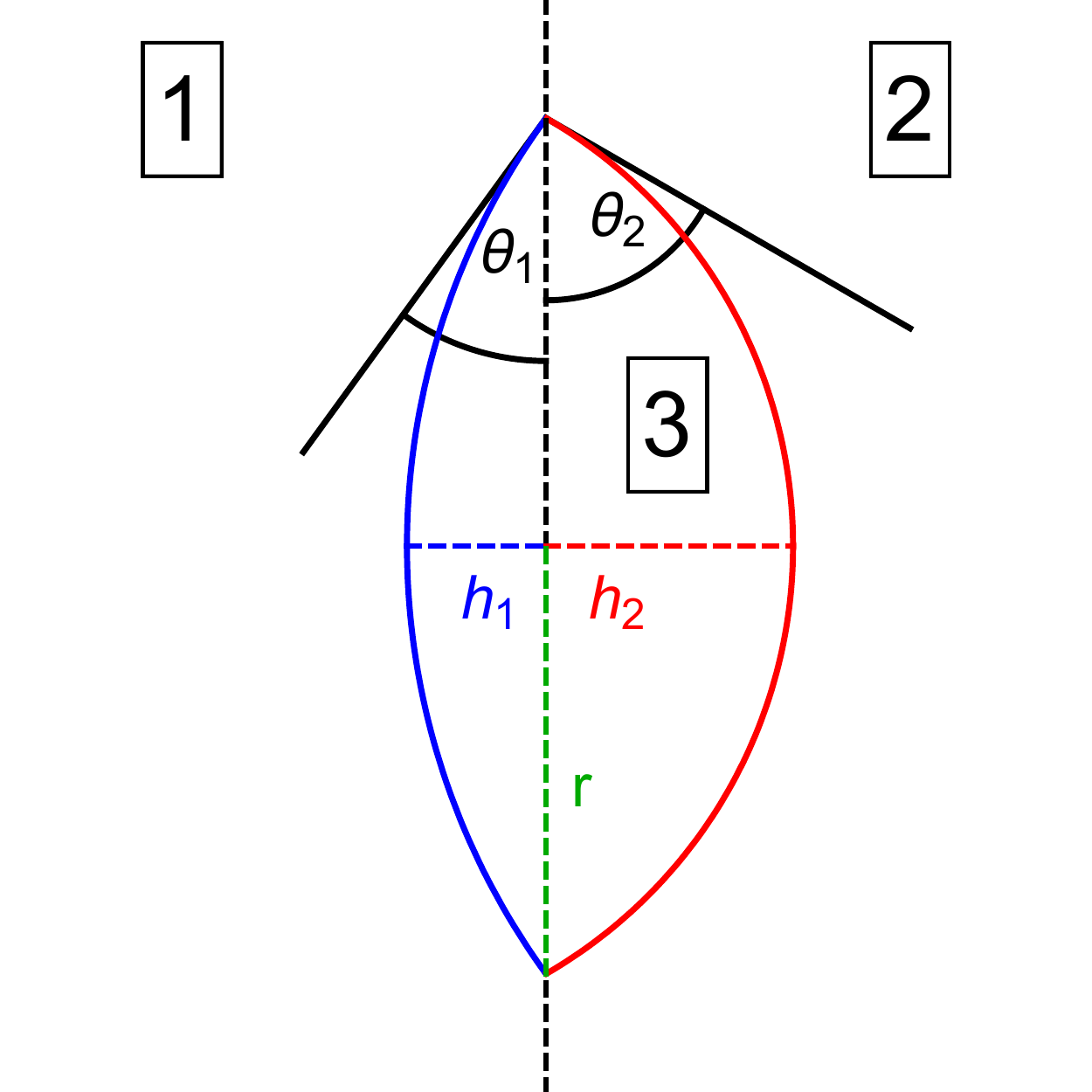}
 \caption{Two-dimensional projection of a bubble of the stable phase (region 3) that has formed on the boundary of a bulk metastable phase (region 1) and an impurity in the bulk (region 2). The characteristic size of the impurity is assumed to be much larger than that of the bubble such that the boundary between regions 1 and 2 can be treated as a flat plane. This picture is symmetric about the horizontal (blue-red) line in the additional dimensions that aren't shown.}
  \label{fig:wet}
\end{figure}
\begin{figure}
  \centering
  \subfloat[]{\label{fig:wetB}\includegraphics[width=0.4\textwidth]{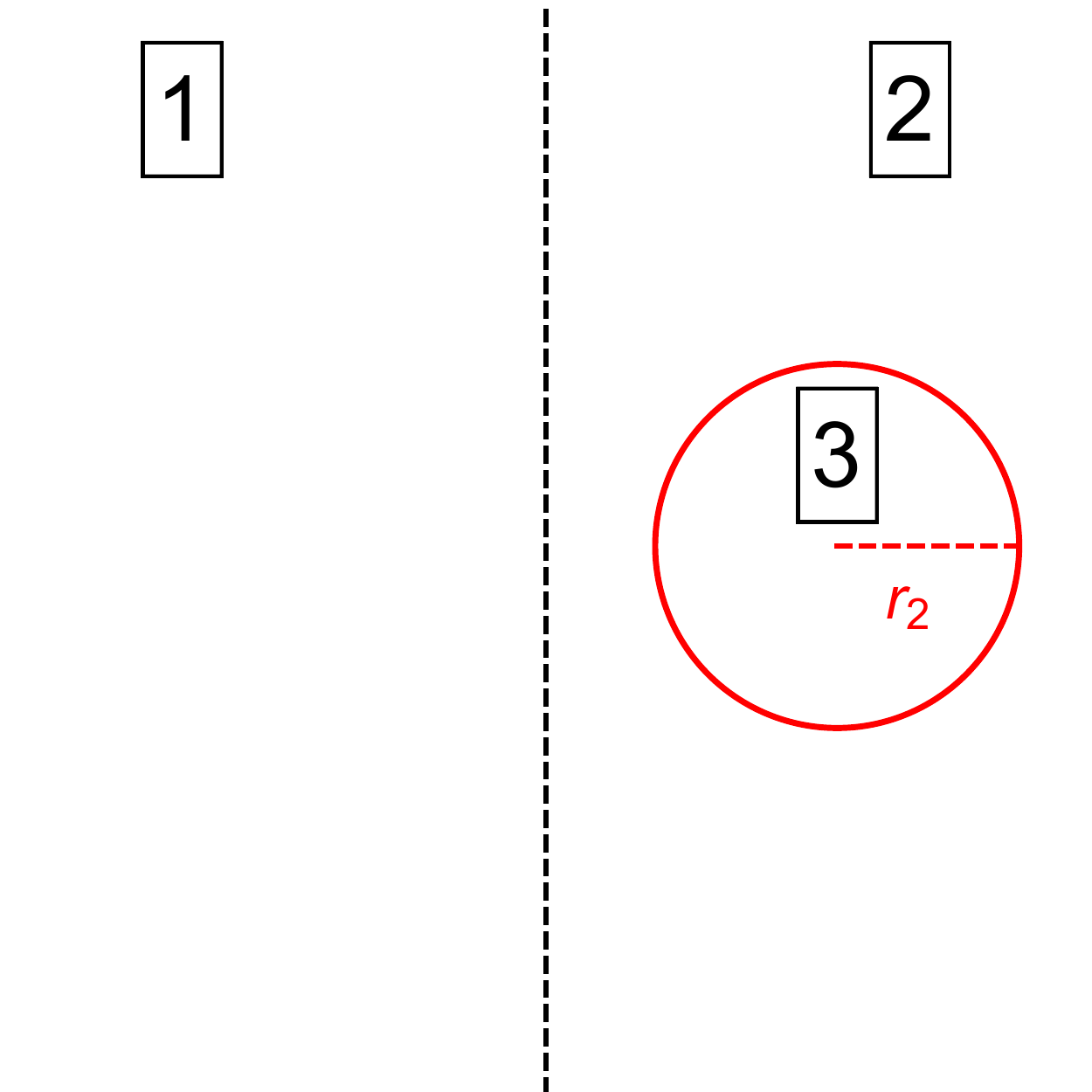}}\,
\subfloat[]{\label{fig:wetA}\includegraphics[width=0.4\textwidth]{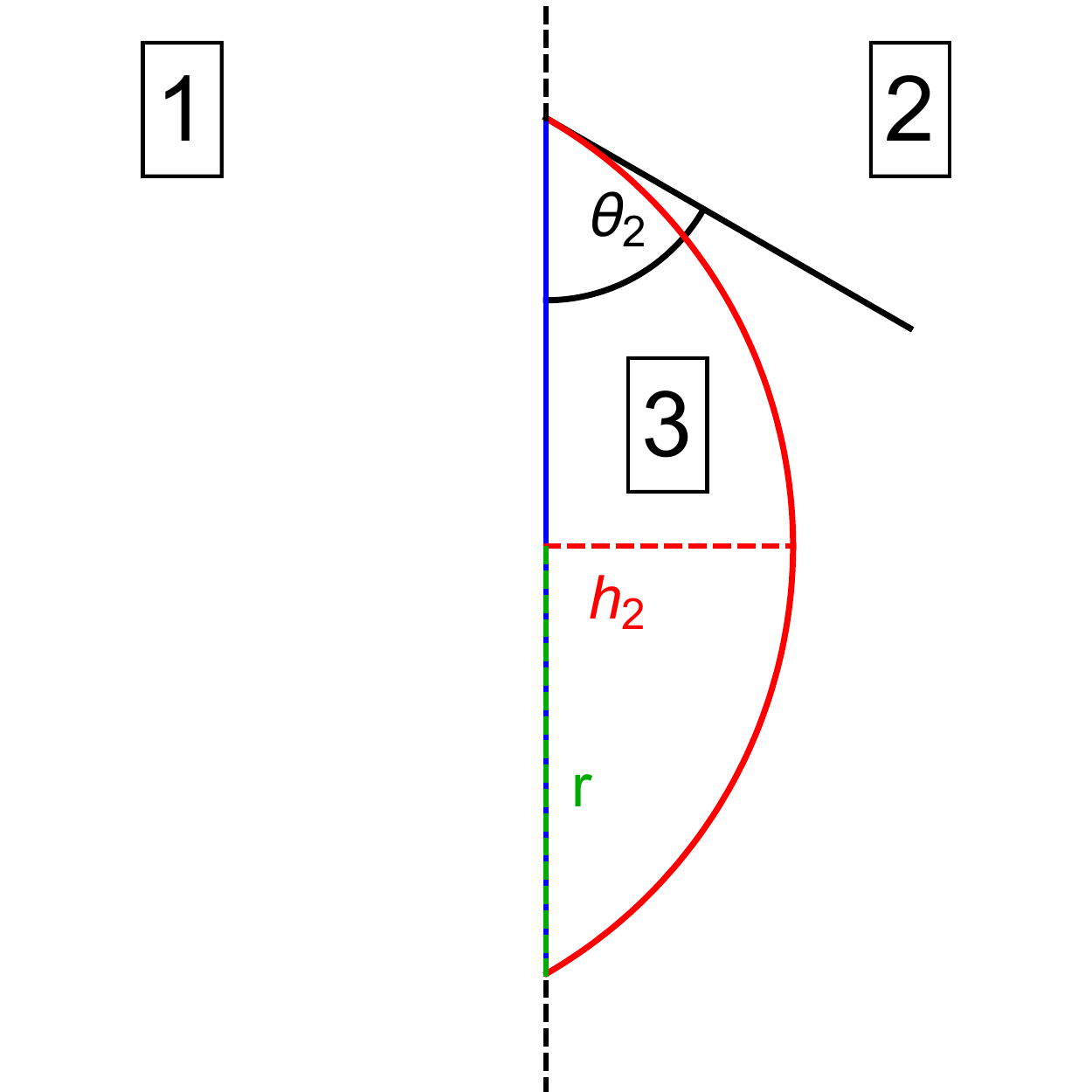}}
 \caption{Special cases of the situation presented in Fig.~\ref{fig:wet}. In Fig.~\ref{fig:wetB}, a full sphere forms within the impurity. While in Fig.~\ref{fig:wetA}, a ``half-bubble'' forms on the interface of regions 1 and 2, but only on the side of the impurity.  Note that the distance between the bubble and the interface between phases 1 and 2 is not necessarily to scale in Fig.~\ref{fig:wetB}. See the text for more details on these special cases.}
  \label{fig:2wet}
\end{figure}

In the case of heterogeneous nucleation due to the presence of an impurity in the bulk, the bubble is no longer a full sphere. Instead, the generalized bubble has a lenticular shape formed by joining two spherical segments as shown in Fig.~\ref{fig:wet}; see Fig.~\ref{fig:2wet} for a couple of special cases. It is this decreased volume of the bubble in the heterogeneous case that leads to the smaller action, and thus in principle a faster decay rate. We assume in this analysis that the characteristic size of the impurity is much larger than that of the bubble such that the boundary between regions 1 and 2 can be treated as a flat plane. It would be interesting to relax this assumption to study the cases where the impurity is a string or monopole of comparable size. Fig.~\ref{fig:wet} is a two-dimensional projection of the bubble, and is symmetric about the horizontal (blue-red) line in the additional (Euclidean) dimensions that aren't shown. 

The Euclidean action in the heterogeneous case can be written as follows,
\begin{align}
\label{eq:newS}
S_{E, \theta} &= S_{E,1}\, f\left(\theta_1\right) + S_{E,2}\, f\left(\theta_2\right) - \frac{4}{3} \pi r^3 \left(S_{12} - S_{12}^{\star}\right), \\
S_{E,i} &= - \frac{\pi^2}{2} r_i^4 \epsilon_{i3} + 2 \pi^2 r_i^3 S_{i3}, \nn \\
f\left(\theta\right) &= \frac{1}{\pi} \left[\theta - \sin\theta \cos\theta \left(1+ \frac{2}{3} \sin^2\theta\right)\right], \nn \\
S_{12}^{\star} &= S_{13} \cos \theta_1 + S_{23} \cos \theta_2, \nn
\end{align}
where the subscript $\theta$ labels the heterogeneous case. In the preceding equations, $\epsilon_{ij}$ and $S_{ij}$ are the energy density difference and the surface tension between phases $i$ and $j$ respectively. As in Fig.~\ref{fig:wet}, regions 1, 2, and 3 correspond to the metastable phase, the impurity, and the stable phase respectively. $S_i$ is action associated with a bubble that forms completely in region $i$, see Eq.~\eqref{eq:S}. The form of the first two terms on the right-hand side of Eq.~\eqref{eq:newS} come from cutting off the angular integration of the spheres at $\theta_{1,2}$. $f\left(\theta\right)$ is the fractional volume of a 4-sphere capped at angle $\theta$ rather than $\pi$ (i.e. $f\left(\pi\right) = 1$, $f\left(\pi/2\right) = 1/2$, etc.). The relevant geometry results can be found in e.g.~\cite{li2011concise}. The last term on the right-hand side of Eq.~\eqref{eq:newS} comes from change in the action caused by the appearance of the stable phase, which eliminates some of the hyperplanar boundary between the metastable phase and the impurity.\footnote{There are an infinite number of ways to return to the homogeneous action from the heterogeneous case. A particularly simple way is to send $\theta_1 \to \pi$; $\theta_2, r \to 0$.} Lastly, $S_{12}^{\star}$ is the value $S_{12}$ takes when a bubble of critical size forms on the interface between the metastable phase and the impurity. As can be seen from Fig.~\ref{fig:wet}, only three of five length/angle parameters in Eq.~\eqref{eq:newS} are independent. One way to write the constraints is,
\begin{equation}
\label{eq:rcon}
r = r_1 \sin \theta_1 = r_2 \sin \theta_2.
\end{equation}
From this, it immediately follows that,
\begin{equation}
\label{eq:h}
h_{1,2} = r_{1,2} \left(1 - \cos \theta_{1,2}\right).
\end{equation}

As was done in the homogeneous case, we now look for the critical points of the heterogeneous action, using $r$, $h_1$, and $h_2$ as our free parameters. However, in what follows we will switch between various combinations of $r,\, h_i,\, r_i$, and $\theta_i$ to write the equations in their simplest form. We find 12 unique critical points, five of which can be eliminated by requiring $r$, $h_1$, and $h_2$ to be real and non-negative. The first derivatives of $S_{E,\theta}$ are,
\begin{align}
\label{eq:1stD}
\frac{d S_{E,\theta}}{d r} &= S_{E3,1}\, g\left(\theta_1\right) + S_{E3,2}\, g\left(\theta_2\right) - 4 \pi r^2 \left(S_{12} - S_{12}^{\star}\right), \\
\frac{d S_{E,\theta}}{d h_i} &= \pi r_i^2 \left(- r_i \epsilon_{i3} + 3 S_{13}\right) h\left(\theta_i\right), \nn \\
S_{E3,i} &= -\frac{4}{3} \pi r_i^3 \epsilon_{i3} + 4 \pi r_i^2 S_{i3}, \nn \\
g\left(\theta\right) &= \frac{1}{4} \cot\left(\frac{\theta}{2}\right)\left(6\theta - 3 \sin\left(\theta\right) - 3\sin\left(2\theta\right) + \sin\left(3\theta\right)\right), \nn \\
h\left(\theta\right) &= \csc^2\left(\frac{\theta}{2}\right)\left(\theta \cos\left(\theta\right) - \frac{3}{4} \sin\left(\theta\right) - \frac{1}{12} \sin\left(3\theta\right)\right). \nn
\end{align}

The simplest critical points to analyze are when a full sphere (or no sphere) forms completely within the bulk of the metastable phase and/or the impurity, $r^{\star} \to 0, h_i^{\star} \to \{0,\, 6 S_{i3} / \epsilon_{i3}$\}. For these cases, the critical action takes simple forms,
\begin{equation}
\label{eq:SEstar1}
S_{E,\theta}^{\star} = \{0,\, S_{E,1}^{\star},\, S_{E,2}^{\star},\, S_{E,1}^{\star} + S_{E,2}^{\star}\}.
\end{equation}
The first solution is the case when no bubble forms, which we aren't interested in. While the second solution is just the homogeneous case. The third solution is a bubble that forms completely within the impurity, as in Fig.~\ref{fig:wetB}. For some range of parameters, this solution will be energetically favorable with respect to the homogeneous case. The same cannot be said for the last of these solutions, which is always energetically disfavored compared to the homogeneous case.

In addition to being energetically favorable with respect to the homogeneous case, for a heterogeneous solution to be of interest, the bubble must grow. The second derivatives of $S_{E,\theta}$, evaluated at the critical values for $r_i$, are,
\begin{align}
\left.\frac{d^2 S_{E,\theta}}{d h_i^2}\right|_{r_i = 0} &= \left.\frac{d^2 S_{E,\theta}}{d r^2}\right|_{r_1 = 0,\, r_2 = 0} = 0, \\
\left.\frac{d^2 S_{E,\theta}}{d h_i^2}\right|_{r_i = \frac{3 S_{i3}}{\epsilon_{i3}}} &= - \frac{9 \pi}{2} \frac{S_{i3}^2}{\epsilon_{i3}} \cos\left(\theta_i\right) \csc^2\left(\frac{\theta_i}{2}\right) h\left(\theta_i\right), \nn \\
\left.\frac{d^2 S_{E,\theta}}{d r^2}\right|_{r_i = \frac{3 S_{i3}}{\epsilon_{i3}},\, r_j = 0} &= - 3 \pi \frac{S_{i3}^2}{\epsilon_{i3}} \cot^2\left(\frac{\theta_i}{2}\right) \left(6 \theta_i - 8 \sin\left(\theta_i\right) + \sin\left(2 \theta_i\right)\right), \nn \\
\left.\frac{d^2 S_{E,\theta}}{d r^2}\right|_{r_1 = \frac{3 S_{13}}{\epsilon_{13}},\, r_2 = \frac{3 S_{23}}{\epsilon_{23}}} &= \left.\frac{d^2 S_{E,\theta}}{d r^2}\right|_{r_1 = \frac{3 S_{13}}{\epsilon_{13}},\, r_2 = 0} + \left.\frac{d^2 S_{E,\theta}}{d r^2}\right|_{r_2 = \frac{3 S_{23}}{\epsilon_{23}},\, r_1 = 0}. \nn
\end{align}
For the third solution of Eq.~\eqref{eq:SEstar1}, the second partial derivative test is ambiguous because one of the eigenvalues is zero, whereas the last solution is found to be a local maximum. However, a numerical study shows that the second and third solutions in Eq.~\eqref{eq:SEstar1} are saddle points of $S_{E,\theta}$ (with respect to $r$, $h_1$, and $h_2$). Thus, there is the possibility of bubble growth after formation. All the cases that are discussed in what follows are also saddle points.

The dynamics of bubble growth in Lorentzian space are easy to picture for these solutions. The bubble grows similarly to how a bubble would grow in the homogeneous case, as described by Coleman~\cite{Coleman:1977py}. The only difference being that for a bubble forming completely within an impurity, the bubble first converts the impurity to true vacuum, before continuing to expand (at the speed of light) out into bulk false vacuum, which is also then converted to true vacuum.

The second most simple case is when a bubble forms at the interface between the metastable phase and the impurity, but only on one side of the interface as in Fig.~\ref{fig:wetA}. We dub this scenario the ``half-bubble.'' An example of such a scenario is briefly discussed in Sec.~\ref{sec:scale}. First consider the case where $h_2 = 0$. The case where $h_1 = 0$ is trivial to obtain from the solution with $h_2 = 0$. The critical sizes in this scenario are,
\begin{equation}
\label{eq:h1r}
\left(h_1^{\star}, r^{\star}\right) = \left(\frac{3\left(S_{13} - S_{12} + S_{23}\right)}{\epsilon_{13}}, \frac{3 \sqrt{S_{13}^2 - \left(S_{12} - S_{23}\right)^2}}{\epsilon_{13}}\right).
\end{equation}
From this, we see that $S_{12} - S_{23} = S_{13} \cos \theta_1$. The critical action in this case is,
\begin{equation}
\label{eq:SEh1r}
S_{E,\theta}^{\star} = S_{E,1}^{\star}\, f\left(\theta_1^{\star}\right),
\end{equation} 
with $\cos \theta_1^{\star} = \left(S_{12} - S_{23}\right) / S_{13}$. To obtain the analogous solution for when a bubble forms on the interface within the impurity, but not in the metastable phase, simply switch the 1's and 2's in Eqs.~\eqref{eq:h1r} and~\eqref{eq:SEh1r}, i.e. $S_{E,\theta}^{\star} = S_{E,2}^{\star}\, f\left(\theta_2^{\star}\right)$ with $\cos \theta_2^{\star} = \left(S_{12} - S_{13}\right) / S_{23}$.

The remaining solution is the most general scenario (in terms of parameters), where a bubbles forms at the interface and on both sides of the interface, see Fig.~\ref{fig:wet}. In this case, it is easier to express the conditions for criticality in terms of $r_i$ and $\theta_i$, rather than $r$ and $h_i$,
\begin{align}
r_i^{\star} &= \frac{3 S_{i3}}{\epsilon_{i3}}, \\
\cos \theta_i^{\star} &= \frac{S_{12} \epsilon_{i3}^2 - \sqrt{S_{j3}^2 \epsilon_{i3}^4 + S_{i3}^2 \epsilon_{j3}^4 + \left(S_{12}^2 - S_{i3}^2 - S_{j3}^2\right) \epsilon_{i3}^2 \epsilon_{j3}^2}}{S_{i3} \left(\epsilon_{i3}^2 - \epsilon_{j3}^2\right)}, \nn
\end{align}
for $i = 1, 2$ and $j = 2, 1$. Note that there is a similar critical point, with a plus sign rather than minus sign in its solution for $\cos \theta_i^{\star}$. However, this leads to an imaginary value for $r$; $\cos \theta_i = S_{12} \epsilon_{i3} + \sqrt{()} \ldots \to r^2 < 0$. Just as in the half-bubble case, $S_{12} = S_{12}^{\star}$, leading to a critical action of the following form,
\begin{equation}
S_{E,\theta}^{\star} =  S_{E,1}^{\star}\, f\left(\theta_1^{\star}\right) + S_{E,2}^{\star}\, f\left(\theta_2^{\star}\right).
\end{equation}
It will prove useful to examine the limit $\epsilon_{j3} \to 0$, in which the angles take the following forms,
\begin{equation}
\cos \theta_i = \frac{S_{12} - S_{j3}}{S_{i3}}, \quad \cos \theta_j = 1.
\end{equation}
These angles are the same as in the half-bubble case, and in fact this limit reduces to the half-bubble case. We now a have way to determine which side of the interface the half-bubble will form on, and that is the side $i$, on which $\epsilon_{i3} \neq 0$.

Lastly, we discuss the pre-exponential factor in the heterogeneous case, $A_{\theta}$. This factor can be thought of as an attempt frequency that takes into account the numbers of sites available for a heterogeneous decay to occur~\cite{doi:10.1021/jp056377e}. Assuming all the impurities are the same, in analogy with the homogeneous case, quantity of interest in the heterogeneous case should the decay rate per number of impurities per unit surface area, $\Gamma / (N_{\text{imp.}} \mathcal{S}_2)$. If this is so, then $A_{\theta}$ should be proportional to $1 / (r^{\star})^3$ by dimensional analysis. Because there is a smaller amount of symmetry in the heterogeneous case, $A_{\theta}$ is typically smaller than $A$, meaning that $S_{E,\theta}$ cannot be arbitrarily smaller than $S_E$ and still increase the decay rate.\footnote{More precisely, the pre-exponential factor in the probability (rather than decay rate) is typically smaller in the heterogeneous case.} There is some minimum decrease in the action needed to overcome the suppression of the pre-exponential factor to obtain an enhanced decay rate.

\section{Thin-Wall Examples}
\label{sec:ex}
In the previous section, we showed that the action associated with the decay rate of a metastable phase in the presence of an impurity has several additional parameters with respect to the homogeneous case. The question then becomes, how does one calculate those parameters from a microscopic theory. As it is the simplest case, we will mostly focus on the scenario where the bubble forms completely within the bulk of the impurity, Fig.~\ref{fig:wetB}. In particular, we give two examples in the thin-wall approximation that show how to calculate $S_{23}$ and $\epsilon_{23}$ in these examples. In both examples, we give an expression for $S_{E,\theta}$, but save the analysis of the heterogeneous decay rate for Sec.~\ref{sec:H}. We hope that these examples will serve a guide to those who wish the compute the heterogeneous decay rate other models. 

Before we get into the examples, we give a quick review of the thin-wall model of~\cite{Coleman:1977py}. The Euclidean Lagrangian is,
\begin{equation}
\label{eq:Lho}
\mathcal{L}_E = \frac{1}{2}\left(\partial \phi_1\right)^2 +  \frac{\lambda_1}{8}\left(\phi_1^2 - a_1^2\right)^2 + \frac{\epsilon_{13}}{2 a_1}\left(\phi_1 - a_1\right),
\end{equation}
where the subscripts are for later convenience. The solution, $\phi_1(\rho)$, is invariant under four-dimensional Euclidean rotations, with $\rho = \sqrt{t_E^2 + \left|\mathbf{x}\right|^2}$ being the 4-$d$ radial coordinate. The thin-wall approximation is $\lambda_1 a_1^4 \gg \epsilon_{13}$, such that the energy difference between the two vacua is small, and that the Lagrangian has a $Z_2$ symmetry to an excellent approximation. In the thin-wall approximation, $\phi_1$ can be expressed as,
\begin{equation}
\label{eq:phi1}
\phi_1\left(\rho\right) = 
\begin{cases} 
-a_1 & \quad \rho \ll r_1 \\ 
a_1 \tanh\left(\frac{\mu_1}{2} \left(\rho - r_1\right)\right) & \quad \rho \approx r_1 \\
a_1 & \quad \rho \gg r_1
\end{cases},
\end{equation}
with $\mu_1 = \sqrt{\lambda_1} a_1$ and $r_1$ being the radius of the bubble (in the homogeneous case). From Eq.~\eqref{eq:phi1}, we see that the field configuration deep within the bubble is the true vacuum, and that far outside the bubble is the false vacuum. While near the wall, Eq.~\eqref{eq:phi1} shows that the field configuration is a soliton, which smoothly interpolates between the two vacua. This solution for $\phi_1$ yields a Euclidean action of the form of Eq.~\eqref{eq:S} up to corrections of order $O\left(\mu_1^{-2} r_1^{-2}\right)$, which are small in the thin-wall approximation. With the solution for $\phi_1$, the surface tension of the bubble can be calculated in the usual way,
\begin{equation}
\label{eq:s13}
S_{13} = \int_0^{\infty}\! d\rho\, \mathcal{L}_E = \frac{2}{3} \sqrt{\lambda_1} a_1^3 = \frac{2}{3} \frac{\mu_1^3}{\lambda_1}.
\end{equation}

There are a couple of changes in the heterogeneous case. One difference is that, in general, the impurity breaks the O$(4)$ symmetry of the homogeneous solution. Non-O$(4)$ symmetric tunneling solutions have been studied by Refs.~\cite{Hiscock:1987hn, Berezin:1987ea, Arnold:1989cq, Berezin:1990qs} in the context of black holes as the seeds of the bubble nucleation. In addition, it is well known that the tunneling solution at finite temperature possess an O$(3)$ symmetry, rather than the full O$(4)$ symmetry~\cite{Sher:1988mj, Rose:2015lna}. 

Another notable feature of the heterogeneous case is that in general there are additional field configurations for $\phi_1$, beyond those listed in Eq.~\eqref{eq:phi1}. These regimes are due to the impurity, and can be found by solving the equation of motion for $\phi_1$ including the $\phi_1$ field's interactions with the impurity. Regrettably, we do not know of any such analytic solutions for the examples we discuss in what follows. However, some progress has recently been made in this direction, at least in the case of radiatively generated potentials~\cite{Garbrecht:2015cla, Garbrecht:2015yza}. To proceed, we do not consider the full parameter space of these examples, but rather we make choices for some subset of parameters that allows to use the homogeneous solution for $\phi_1$ in our calculation of the heterogeneous decay rate. In any case, $S_{23}$ is given by an expression analogous to Eq.~\eqref{eq:s13} except that the Euclidean Lagrangian now also includes $\phi_1$'s interactions with the impurity. Typically, $\epsilon_{23}$ can simply be read off of the Lagrangian that includes $\phi_1$'s interactions with the impurities, as is done for $\epsilon_{13}$ in the homogeneous case.

\subsection{Scalars}
\label{sec:scale}
As a first example, consider adding a second scalar field to the thin-wall model of~\cite{Coleman:1977py}. The Lagrangian is,
\begin{align}
\mathcal{L} &= \frac{1}{2}\left(\partial \phi_1\right)^2 + \frac{1}{2}\left(\partial \phi_2\right)^2 - U\left(\phi_1,\, \phi_2\right), \\
U &= \frac{\lambda_1}{8}\left(\phi_1^2 - a_1^2\right)^2 + \frac{\lambda_2}{8}\left(\phi_2^2 - a_2^2\right)^2 + \frac{\lambda_3}{8}\left(\phi_1^2 - a_1^2\right) a_2 \left(\phi_2 - a_2\right) + \frac{\epsilon_{13}}{2a_1} \left(\phi_1 - a_1\right) \nn.
\end{align}
Here $\phi_1$ is the bounce solution that governs the tunneling rate, and $\phi_2$ is the additional scalar field. We will assume the field configuration of $\phi_2$ is that of a kink, say, in the $z$-direction, centered about $z_0$, which breaks the O$(4)$ symmetry of the homogeneous solution and provides a preferred direction for tunneling. We will also assume that the surface tension of the $\phi_2$ kink is large enough that the backreaction of $\phi_1$ on $\phi_2$ is negligible, and that it is narrow enough such that the boundary between regions 1 and 2 can be treated as a flat plane. Both of these assumptions can be accommodated by taking $\mu_2 = \sqrt{\lambda_2} a_2$ to be much larger than any other energy scale in the problem.

For $z > z_0$, the $\lambda_3$ interaction has no effect as $\phi_2 \approx a_2$ in this region except for the transition near the kink of $\phi_2$. However, this transition region should be small if the kink appears as a plane to the bubble, which we have assumed. Therefore, the surface tension in this direction, $S_{13}$, is the same as in the homogeneous case. For $z < z_0$, the effect of the $\lambda_3$ interaction is to shift the mass term of the first scalar field, $\mu_1^2 \to \mu_1^2 + \lambda_3 a_2^2$, where again, we have neglected the width of the kink of $\phi_2$.  The surface tension of the wall between the bubble of stable phase and the impurity is then given by $S_{13}$, but with the aforementioned shifted mass,
\begin{equation}
S_{23} \approx \frac{2}{3 \lambda_1} \left(\mu_1^2 + \lambda_3 a_2^2\right)^{3/2} = S_{13}\left(1 + \frac{\lambda_3 a_2^2}{\lambda_1 a_1^2}\right)^{3/2}.
\end{equation}
The difference between the energy density of the two vacua now depends on the shifted mass as well,
\begin{equation}
\epsilon_{23} = \epsilon_{13} \sqrt{1 + \frac{\lambda_3 a_2^2}{\lambda_1 a_1^2}}.
\end{equation}

From this, we see that for $\lambda_3 > 0$, there is no possibility of an enhanced tunneling rate for a bubble forming completely within the impurity, as the action increases in this scenario. However, when $\lambda_3 < 0$ and $\left|\lambda_3\right| a_2^2 < \lambda_1 a_1^2$, the heterogeneous action for a bubble to form within the impurity is given by,
\begin{equation}
\label{eq:scalar}
S_{E,\theta} = S_E \left(1 - \frac{\left|\lambda_3\right| a_2^2}{\lambda_1 a_1^2}\right)^{9/2},
\end{equation}
which does lead to the possibility of an enhanced decay rate contingent upon the change in the pre-exponential factor in the heterogeneous case relative to the homogeneous case.

On the other hand, for $\lambda_3 < 0$ and $\left|\lambda_3\right| a_2^2 > \lambda_1 a_1^2$, there is a unique vacuum at $\phi_1 = 0$ for $ z < z_0$ (ignoring the tiny width of the $\phi_2$ kink) with an energy density of $\lambda_1 a_1^4 / 8$ up to corrections of order $\epsilon_{13}$.  While for $ z > z_0$, the two vacua of the homogeneous case are still present at $\phi_1 = \pm a_1$. Comparing energy densities, we see that $\epsilon_{23} = O(\lambda_1 a_1^4) \gg \epsilon_{13}$. Based on this comparison and the discussion of Sec.~\ref{sec:D}, the picture for this scenario is given by Fig.~\ref{fig:wetA}. However, this comparison also calls into question whether the thin-wall approximation is still valid, as $\epsilon_{23} \sim S_{13}$. For this reason, we do not consider this range of parameters further in this work. 

\subsection{Fermions}
\label{sec:ferm}
Consider adding a fermion to the standard thin-wall example of~\cite{Coleman:1977py}. The fermion has a Yukawa interaction with the scalar,
\begin{equation}
U \supset y \bar{\psi} \psi \phi_1 .
\end{equation}
Assume that in some regions of space, there is a non-zero fermion number density, $N_{\psi}\left(\left|\mathbf{x}\right|\right) = \langle \bar{\psi} \psi \rangle \neq 0$. These fermions might condense into composite objects, like in QCD, or there may simply be a non-zero number of elementary fermions. In either case, we will assume that the scalar interacts with this object through the Yukawa term.

The difference between the homogeneous and heterogeneous cases is the addition of the Yukawa interaction, so $\phi_1$ will be required to tunnel through the finite density of fermions in some regions of space. The surface tension between the metastable phase and the impurity should be given by,
\begin{equation}
S_{23} = S_{13} + y N_\psi \int_{0}^{L}\! d\rho\, \phi_1(\rho),
\end{equation}
if the following two assumptions are satisfied: (1) $L \gg r_{1,2}$, in order to satisfy the assumption that the boundary between the metastable phase and the impurity appears as a plane to the bubble, and (2) the fermion density is constant, $N_{\psi}\left(\left|\mathbf{x}\right|\right) = N_{\psi}$, in the regions of space where it is non-zero. Experience from the real world suggests that $L \gsim N_{\psi}^{-1/3}$; the majority of the nuclei in the universe are light, and they are all at least as big as a proton, which has a size $\sim N_{\text{QCD}}^{-1/3}$. Taking $L \gsim N_{\psi}^{-1/3}$, the planar assumption suggests that $N_{\psi}$ is small enough such that the homogeneous solution for $\phi_1$ is not significantly perturbed, which yields
\begin{equation}
\label{eq:s23f}
S_{23} - S_{13} \approx  \frac{2}{\sqrt{\lambda_1}} y N_{\psi} \ln\left(\cosh\left(\frac{\sqrt{\lambda_1}}{2} a_1 N_{\psi}^{-1/3}\right)\right).
\end{equation}
Combing the planar assumption ($L \gg r_1$) with the thin-wall approximation ($\lambda_1 a_1^4 \gg \epsilon_{13}$), we find $\sqrt{\lambda_1} a_1 L \gg \lambda_1 a_1^4 / \epsilon_{13} \gg 1$. Thus, Eq.~\eqref{eq:s23f} can be rewritten as,
\begin{equation}
S_{23} - S_{13} \approx y N_{\psi}^{2/3} a_1 - O(N_{\psi}).
\end{equation}

The difference in the energy density between the false and true vacua in the heterogeneous case can be read off of the Lagrangian,
\begin{equation}
\epsilon_{23} = \epsilon_{13} + 2 y N_{\psi} a_1.
\end{equation}
For a bubble that forms completely within the impurity, the critical Euclidean action in the heterogeneous case is given by,
\begin{equation}
S_{E,\theta} = \frac{27 \pi^2}{2}\frac{\left(S_{13} + y N_{\psi}^{2/3} a_1\right)^4}{\left(\epsilon_{13} + 2 y N_{\psi} a_1\right)^3} \approx S_E \left(1 + \frac{4y N_{\psi}^{2/3} a_1}{S_{13}} - O(N_{\psi})\right).
\end{equation}
For $y < 0$, the preceding equation gives a suppressed action, and thus the possibility of an enhanced decay rate. However, since $N_{\psi}$ has been assumed to be small, its effect on the action will be small in this example regardless of the sign of $y$. Note that to decrease the action with these assumptions, $y$ must be negative whether $-a$ or $a$ ($\epsilon_{13} \to -\epsilon_{13}$ in Eq.~\eqref{eq:Lho}) is chosen as the true vacuum in the homogeneous case. 

\section{Application to the Higgs Vacuum}
\label{sec:H}
In this section, we apply the formalism developed in the previous two sections to the case of the Higgs vacuum with baryonic matter as impurities, as well as investigate generic BSM physics. 

We start by briefly reviewing the homogeneous vacuum decay rate and probability. See Refs.~\cite{DiLuzio:2015iua, Isidori:2001bm} for more details. In the case of the Higgs vacuum, the decay rate per unit volume in the SM is given by
\begin{equation}
\label{eq:Hdecay}
\frac{\Gamma}{\mathcal{V}_3} \approx \frac{1}{R^4}\, e^{-S\left(\Lambda_B\right)}.
\end{equation}
Since quantum corrections enter the potential logarithmically, one must go to large field values to determine if there is an instability, $h \gg v$. With this approximation, the bounce action in Eq.~\eqref{eq:Hdecay} can be obtained using the tree level potential,
\begin{equation}
S\left(\Lambda_B\right) = \frac{8 \pi^2}{3\left| \lambda\left(\Lambda_B\right)\right|}.
\end{equation}
This action is classically scale invariant, and the size of the bounce, $R$, is arbitrary. Quantum corrections break scale invariance, contributing to action as $\Delta S \sim \log\left(R\, \Lambda_B\right)$. Choosing $\Lambda_B \sim R^{-1}$ minimizes these corrections, and resolves the implicit ambiguity in choosing which $\overline{\text{MS}}$ scale to evaluate $\lambda$ at. As always, one is interested in the bounce that maximizes the decay rate. With this choice of scale, the maximization condition is in practice given by $\beta_{\lambda}(\Lambda_B) = 0$, where $\Lambda_B \approx 2.0 \times 10^{17}$ GeV in the SM~\cite{Buttazzo:2013uya}. Eq.~\eqref{eq:Hdecay} can be integrated over the past light cone of the universe to give the probability that the Higgs vacuum decayed at some point in the past light cone of the universe, 
\begin{equation}
P_0 \approx \left(\frac{\Lambda_B^4}{H_0^4}\right) e^{-S\left(\Lambda_B\right)},
\end{equation}
where $H_0$ is the Hubble constant. Numerically, we find $P_0 \sim 10^{-741}$, compare against the left panel of Fig.~7 of Ref.~\cite{Buttazzo:2013uya}.

After clarifying the assumptions that go into the calculation, we proceed to investigate the case of baryonic matter (i.e. stars) in the universe acting as impurities to seed the decay of the Higgs vacuum. 

First of all, it should be noted that the examples of the previous section relied on the thin-wall approximation, which is not valid for the SM. One way around this problem, which was the approach of~\cite{Burda:2015yfa}, is to include additional, Planck-scale suppressed operators (for example: $(H^{\dagger}H)^3 / M_{\text{pl}}^2$) to modify the Higgs potential such that it does satisfy the conditions for the thin-wall approximation to be used.  We do not include any such operators in our calculation, as we believe our results will not qualitatively be affected by this approximation, and hope that it will be self-evident as to why we think so in what follows. Furthermore, if such operators were included, it may be difficult to separate the effects of the impurity from the effects of the operators.

In addition, it was shown in Sec.~\ref{sec:ferm} that (at least for a bubble forming completely within the impurity, and $N_{\psi}^{2/3} a_1 \ll S_{13}$) the Yukawa coupling of the fermions to the scalar field must be negative to actually cause a decrease in the action. This is also not the case in the SM, and there are at least two ways out of it. The first is to simply choose a more complicated critical point of the heterogeneous action. With a larger set of parameters, it will allow for a decreased action with a positive Yukawa coupling. A second choice is to add an operator (or two) to the theory of the form, $(H^{\dagger} H)(\bar{q}_1 d_1 H)$ or $(H^{\dagger} H)(\bar{q}_1 u_1 \tilde{H})$, where the subscripts are generational indices. These operators can make the effective Yukawa couplings of the up- and down-quarks negative, while keeping their masses positive. The first generation Yukawa coupling also need to be kept small enough in magnitude so as to not affect the stability analysis of the homogeneous case, which is possible to do with this setup. Such a scenario was explored for the top-quark in~\cite{Hedri:2013wea}. 

With the caveats out of the way, we now turn to the analysis. The nuclear matter in stars has a typical density of $N_b \approx \Lambda_{\text{QCD}}^3 \approx (1\, \text{GeV})^3$, whereas the Higgs bounce solution has a characteristic scale $\Lambda_B$. From the fermion example in Sec.~\ref{sec:ferm}, we see that baryonic matter gives a tiny correction factor to the exponent of the decay rate, 
\begin{equation}
\frac{S_{E,\theta}}{S_E} - 1 = O\left(\frac{y_{u,d}  \Lambda_{\text{QCD}}^2}{\Lambda_B^2}\right),
\end{equation} 
with $y$ being the Yukawa coupling of a light quark commonly found in nuclear matter. However, even if it was the top-quark Yukawa, this exponential correction factor would still for all intents and purposes be one as there is such a large disparity between $\Lambda_{\text{QCD}}$ and $\Lambda_B$.

It seems fair to ask, since the correction to the exponent of the decay rate is so small, is the bubble formation process even sensitive to the presence of baryons. If the answer is no, then the probability that the Higgs vacuum decayed in the past light cone of the universe, $P$, is the same tiny number as in the homogeneous case, $P = P_0$. On the other hand, if the answer is yes, then we need to calculate the pre-exponential factor. Based on the analysis on Sec.~\ref{sec:D}, we find, 
\begin{equation}
P = \left(\frac{n_b}{H_0^3}\right) \left(\frac{\Lambda_B^3}{\Lambda_{\text{QCD}}^2 H_0}\right) e^{-S\left(\Lambda_B\right)}.
\end{equation}
The first term in parentheses estimates the number of baryons in the visible universe. An equally good expression is $\Omega_b \rho_{\text{crit}} H_0^{-4}$. Both quantities give the same order of magnitude for the number of baryons in the visible universe, $N_{\text{imp.}} \approx 10^{77}$. All cosmological parameters were taken from~\cite{Ade:2013zuv}. The denominator of the second term in parentheses is an estimate of effective surface area of a proton, with $\Lambda_{\text{QCD}}^{-1}$ being the spatial radius of a proton. The factor of $H_0$ in the second term appears because the proton is stable. We have assumed a hyper-cylindrical shape for the proton (in spacetime), as we believe it gives a better approximation of the pre-exponential factor. The conclusions regarding the exponent of the rate decay are unaffected by this assumption on the shape of a proton. By inspecting the following ratio,
\begin{equation}
\frac{P}{P_0} = \left(\frac{n_b}{H_0^3}\right) \left(\frac{H_0^3}{\Lambda_{\text{QCD}}^2\Lambda_B}\right) \approx \left(10^{77}\right) \left(10^{-143}\right) \ll 1,
\end{equation}
we see that the pre-exponential factor is much smaller in the heterogeneous case than in the homogeneous case, leading to a subdominant heterogeneous decay rate. On that note, it's worth mentioning that Ref.~\cite{Cheung:2013sxa} found that Hawking radiation does not classically catalyze Higgs vacuum decay in the SM.

Even though impurities from within the SM do not catalyze Higgs vacuum decay, it may be possible for impurities from BSM physics to do just that. The potential enhancement of the decay rate hinges on both the scale and the density of the new physics. In particular, they must both be large, relatively speaking. First of all, the stability analysis in the homogeneous SM assumes that there is no new physics up to the Planck scale. If the BSM scale is similar to or less than the instability scale in the SM, which is rather large itself ($\Lambda_I \sim 10^{11}$ GeV, see~\cite{Buttazzo:2013uya} for more information), the BSM physics may well stabilize the Higgs potential. In any case, a stability analysis would have to be performed. Assuming the scale of the BSM physics is higher than $\Lambda_I$ avoids this issue and is more in line with the spirit of the homogeneous SM analysis. Additionally, the results of the examples investigated in Sec.~\ref{sec:ex} show that the decrease of the heterogeneous action relative to the homogeneous case depends on the ratio, $\Lambda_{\text{BSM}}^2 / \Lambda_B^2$. Based on the scalar example, Eq.~\eqref{eq:scalar}, GUT scale new physics with $\left|\lambda_3\right| a_2^2 \sim \left(10^{15}\, \text{GeV}\right)^2$, and the $\Lambda_B^2$ estimated as $\lambda_1 a_1^2 \sim \left(10^{17}\, \text{GeV}\right)^2$, should cause a $4-5\%$ decrease in the heterogeneous action if their interactions with the Higgs field are mediated by a scalar field.\footnote{There is an upper bound on the masses of right-handed neutrinos operating through a see-saw mechanism of $\sim 10^{13-14}$ GeV from the requirement that the lifetime of the EW vacuum is longer than the age of the universe~\cite{EliasMiro:2011aa}. With this bound, the potential for right-handed neutrinos to affect the heterogeneous decay rate is rather small. On the other hand, the EW vacuum can be stabilized by a scalar threshold effect, as shown in~\cite{EliasMiro:2012ay}. The combination of these two effect was recently investigated in Ref.~\cite{Ng:2015eia} with the conclusion that a scalar threshold can stabilize the EW vacuum even if the theory contains right-handed neutrinos.} The decrease in action continues as $\Lambda_{\text{BSM}}$ approaches $\Lambda_B$, until $\Lambda_B$ is reached at which point it's not clear if the thin-wall approximation still holds for the scalar example of Sec.~\ref{sec:scale}. The pre-exponential factor is governed by the density of impurities. In order for a modest decrease in the action to lead to an enhanced decay rate, as is the case for GUT scale physics, the density of impurities large enough such that the pre-exponential factor does not suppress the heterogeneous decay rate.

\section{Discussion and Conclusions}
It was mentioned in the introduction that in statistical mechanics systems there are typically two ways to induce a phase transition, either through an impurity in the bulk or through a boundary of the system. Until now, we focused entirely on impurities in the bulk of the metastable phase. The reason is that a spatial boundary would need to be within our past light cone in order for it to affect the probability that the universe would have already decayed by now, rather than just the decay rate itself. It seems likely that if such a boundary was within the past light cone of the visible universe, it would have already been detected, which is why this scenario is not given much consideration. If there is a spatial boundary of the universe outside of the light cone of the visible universe, the boundary would affect the lifetime of the universe, but not the probability that it has already decayed. Even though bubbles of true vacuum may be forming on this boundary at a faster rate than they would in the bulk, they would not have had enough time to reach our visible universe. 

This raises the question of what should be considered a spatial boundary. We have in mind a universe that is, say, topologically flat with a finite volume bounded by, say, a 4-sphere. This is different from a universe with a non-trivial global topology.\footnote{For an analysis of the global topology of the universe based on CMB temperature and polarization data see~\cite{Aslanyan:2011zp, Aslanyan:2013lsa}.} Similarly, in extra dimensional theories, typically the extra dimensions have periodic boundary conditions, which would not serve as a spatial boundary. For theories without periodic boundary conditions there would likely a suppression of the pre-exponential factor, originating from making the extra dimension(s) small enough to have presently avoided detection.

As previously mentioned, it would be interesting to relax the assumption that the boundary between the impurity and the metastable phase is a flat plane. This would allow for the study of the cases where the impurity is a string or monopole of comparable size. Another possibility for a future direction would be to generalize this formalism to curved space. 

In this work, we derived the decay rate of an unstable phase of a quantum field theory in the presence of an impurity in the thin-wall approximation. This derivation was based on the how the impurity changes the (flat spacetime) geometry relative to case of pure false vacuum. Two examples were given that show how to estimate some of the additional parameters that enter into this heterogeneous decay rate. This formalism was then applied to the Higgs vacuum of the SM, where baryonic matter (stars) acts as an impurity in the electroweak (metastable) Higgs vacuum. We showed that the heterogeneous decay rate is suppressed with respect to the homogeneous case, which is to say that the conclusions drawn from the homogeneous case are not modified by the inclusion of baryonic matter in the calculation. On the other hand, we confirmed that BSM physics with a characteristic scale comparable to $\Lambda_B$, can in principle lead to an enhanced decay rate contingent upon the change in the pre-exponential factor in the heterogeneous case relative to the homogeneous case, which is governed by the density of impurities in the false vacuum.


\begin{acknowledgments}
This work was supported in part by the MIUR-FIRB under grant no.\! RBFR12H1MW (CM), and the U.S. Department of Energy under grant no.\! DE-SC0009919 (BG).  CM thanks Riccardo Barbieri, John McGreevy, Yasunori Nomura, Alessandro Strumia, and Enrico Trincherini for helpful conversations. In addition, CM is grateful for the hospitality of the UCSD and UCB physics departments, where parts of this work were completed. CM also thanks the LNBL theory group and the SCIPP for the opportunities to present preliminary versions of this work. 
\end{acknowledgments}

\bibliography{MSwB2}
\bibliographystyle{jhep}

\end{document}